\documentclass[letterpaper]{article}
\usepackage{aaai25} 
\usepackage{times} 
\usepackage{helvet} 
\usepackage{courier} 
\usepackage[hyphens]{url} 
\usepackage{graphicx} 
\usepackage{graphicx}  
\usepackage{natbib}  
\usepackage{caption}  
\usepackage{xcolor}
\usepackage{algorithm}
\usepackage{algorithmic}

\usepackage{newfloat}
\usepackage{listings}
\usepackage{enumitem}
\usepackage{makecell}
\usepackage{multirow}
\usepackage{multicol}
\usepackage{booktabs}
\usepackage{url}
\usepackage{mathrsfs}
\usepackage{caption}
\usepackage{subfigure}
\usepackage{amsmath}
\usepackage{amsfonts}
\usepackage{dsfont}
\DeclareCaptionStyle{ruled}{labelfont=normalfont,labelsep=colon,strut=off} 
\floatstyle{ruled}
\newfloat{listing}{tb}{lst}{}
\floatname{listing}{Listing}
\title{Beyond Graph Convolution: Multimodal Recommendation \\ with Topology-aware MLPs}
\author{
    Junjie Huang, Jiarui Qin, Yong Yu, Weinan Zhang\thanks{Weinan Zhang is the corresponding author.}\\
}
\affiliations{
    Shanghai Jiao Tong Univeristy, Shanghai, China\\
    \{huangjunjie2019, wnzhang\}@sjtu.edu.cn
}

\begin{document}

\maketitle

\begin{abstract}

Given the large volume of side information from different modalities, multimodal recommender systems have become increasingly vital, as they exploit richer semantic information beyond user-item interactions. Recent works highlight that leveraging Graph Convolutional Networks (GCNs) to explicitly model multimodal item-item relations can significantly enhance recommendation performance. However, due to the inherent over-smoothing issue of GCNs, existing models benefit only from shallow GCNs with limited representation power. This drawback is especially pronounced when facing complex and high-dimensional patterns such as multimodal data, as it requires large-capacity models to accommodate complicated correlations. To this end, in this paper, we investigate bypassing GCNs when modeling multimodal item-item relationship. More specifically, we propose a \underline{T}opology-aware \underline{M}ulti-\underline{L}ayer \underline{P}erceptron (TMLP), which uses MLPs instead of GCNs to model the relationships between items. 
TMLP enhances MLPs with topological pruning to denoise item-item relations and intra (inter)-modality learning to integrate higher-order modality correlations.
Extensive experiments on three real-world datasets verify TMLP's superiority over nine baselines. We also find that by discarding the internal message passing in GCNs, which is sensitive to node connections, TMLP achieves significant improvements in both training efficiency and robustness against existing models.

\end{abstract}
\section{Introduction}\label{sec:intro}
\begin{figure}[t]
    \centering
    \includegraphics[width=\linewidth]{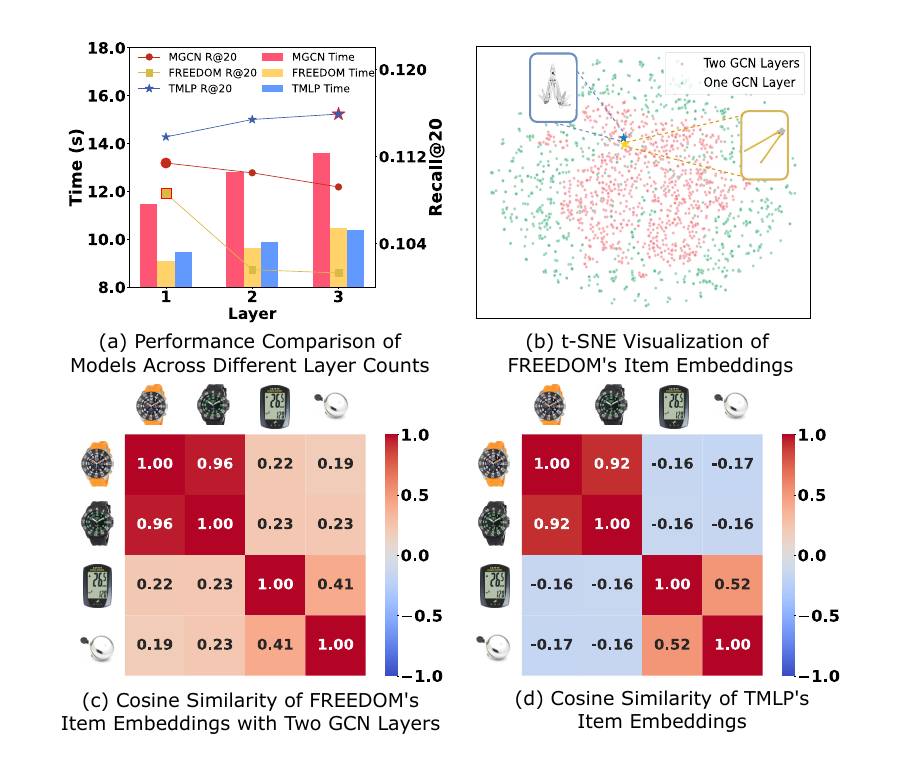}
    \vspace{-7mm}
    \caption{Scalability issues in existing MMRS and motivation for TMLP. Models are well-trained on Amazon Sports.}
    \vspace{-7mm}
    \label{fig:intro}
\end{figure}
Recommender Systems (RS) are indispensable in modern web services, ranging from video streaming platforms to online shopping services~\cite{lin2023can, liu2024mamba4rec, xi2024towards}. Traditional RS~\cite{rendle2012bpr, he2020lightgcn} primarily rely on user-item interactions to learn latent representations. 
However, abundant multimodal information, such as images, videos, and text descriptions related to user and item attributes, is often left unused. This gap motivates the research of Multimodal Recommender Systems (MMRS), which leverage rich semantic multimodal data to improve recommendations~\cite{wei2019mmgcn}. 

The main challenge in MMRS lies in how to effectively incorporate multimodal features within the collaborative filtering (CF) framework~\cite{huang2024comprehensive}. Traditional methods~\cite{he2016vbpr, liu2017deepstyle} enhance item representations by combining multimodal features with item ID embeddings.
Recently, the growing interest in graph-based recommendation methods~\cite{wang2019neural, he2020lightgcn} has led to research utilizing Graph Neural Networks (GNNs) to capture higher-order topological structures from multimodal features.
These studies emphasize that explicitly modeling item-item relationships can significantly improve recommendation performance~\cite{zhang2021mining, yu2023multi}. Existing models typically rely on a single-layer Graph Convolutional Network (GCN) to capture these relations. 
Though achieving remarkable progress, a crucial limitation is its scalability~\cite{yang2022graph}. More specifically, \textbf{due to the \textit{over-smoothing} issue~\cite{chen2020measuring, rusch2023survey}, GCNs are inherently shallow and unable to abide by the scalability law in most deep learning models~\cite{kaplan2020scaling,rosenfeld2019constructive}.}
This limits their capacity to handle complex multimodal data patterns.
In Figure~\ref{fig:intro}, we compare the performance of two state-of-the-art (SOTA) MMRS, FREEDOM~\cite{zhou2023tale} and MGCN~\cite{yu2023multi}, on the Amazon Sports dataset. As GCN layers increase, time per epoch rises, but performance consistently declines. More GCN layers also lead to more concentrated item embeddings, reducing the model's discriminative ability. For instance, in FREEDOM with two GCN layers, unrelated items such as multitools and slingshots power bands (Figure~\ref{fig:intro}(b)), as well as watches and bicycle accessories like bells and computers (Figure~\ref{fig:intro}(c)), are drawn closer together. This implies that, unlike common architectures used in computer vision and natural language lrocessing tasks, given sufficient training data ($n > p$), increasing model capacity does not enhance representation power in GCNs. 
This bottleneck is a significant limitation for current MMRS frameworks.

Consequently, we argue that the complexity of multimodal data and the intricacies of cross-modality correlations may be too challenging for models with limited representation power. 
In this paper, we investigate the possibility of \textbf{capturing topological structure without graph convolution} by introducing \underline{T}opology-aware \underline{MLP} (TMLP). 
TMLP replaces GCNs' message passing with MLPs, which are better suited for handling high-dimensional, complex multimodal data.
By sidestepping the message-passing mechanism, TMLP also gains higher efficiency. To the best of our knowledge, this is the first attempt in a similar line of work to model multimodal correlations without relying on GCNs. 
However, migrating from GCNs to MLPs in MMRS is not straightforward due to the following challenges:
\begin{itemize}[topsep = 3pt,leftmargin =*]
\item \textbf{(C1)} 
\textit{How to train i.i.d model for non-i.i.d data?}
Vanilla MLPs are designed for \textit{identically and independently distributed} (i.i.d.) data, like images, and are topology-agnostic, making them blind to the rich correlations in non-i.i.d. data, such as graph-structured data. The key challenge is designing training objectives that enable MLPs to effectively incorporate topological information.
\item \textbf{(C2)} 
\textit{How to enhance TMLP's robustness with noisy connections in graph-structured data?}
TMLP learns topological dependencies from injected item relations, but noisy connections can mislead the learning process and degrade performance.
Previous efforts~\cite{zhang2021mining, yu2023multi} derive item relations from modality-specific pre-trained models using a simple weighted sum. However, these representations may suffer from modality incompatibility issues, making the item relations noisy.

\end{itemize}

TMLP addresses these challenges with two key modules.
For \textbf{Challenge 1}, TMLP employs \textit{Intra (Inter)-Modality Learning} (IML) to learn topology-aware item representations by maximizing mutual information between an item's hidden representation and its neighborhood~\cite{du2022learning, du2024disco}.
This is efficiently computed using Mutual Information Neural Estimation (MINE)~\cite{belghazi2018mutual}, which transforms mutual information maximization into minimizing neighborhood alignment loss.
This enables TMLP to learn the proximity of items within the same modality (intra-modality correlations) and the representation of similar items across different modalities (inter-modality correlations). 
For \textbf{Challenge 2}, TMLP uses \textit{Topological Pruning Strategy} (TPS) to filter valuable information from pre-trained multimodal features, enhancing robustness against potential mismatches or misalignments between modalities. 
As shown in Figure~\ref{fig:intro}(d), TMLP improves with increased network depth, maintaining the proximity of similar items while keeping unrelated items apart, preserving strong discriminative ability.
In summary, the contributions of this paper are as follows:
\begin{itemize}[topsep = 3pt,leftmargin =*]
\item We propose TMLP, a scalable, efficient, robust, and high-performing MMRS method that pioneers altering the existing framework for item relationship modeling.
\item We develop TPS, which effectively denoises and extracts valuable information from pre-trained features, enhancing robustness in both original and noise-added scenarios.
\item We propose a unified learning procedure that enables modeling intra (inter)-modality correlations altogether, surpassing the explicit message-passing of GCN-based MMRS, thus free from their over-smoothing limitations.
\item Extensive experiments on three real-world datasets show that TMLP significantly outperforms existing SOTA models, achieving over a 7\% performance gain on Amazon Baby, highlighting the empirical superiority of TMLP.
\end{itemize}
\section{Related Work}

\subsection{Graph Collaborative Filtering}~\label{sec:gcf}
User-item interactions can naturally be represented as a bipartite graph, prompting researchers to use GNNs for extracting user behavior features.
Early GNN-based models, grounded in graph spectral theories, are computationally expensive~\cite{monti2017geometric}. More recent efforts have shifted focus to the spatial domain~\cite{wu2018graph, wang2019neural}.
Specifically, NGCF~\cite{wang2019neural} captures user behavior features by iteratively aggregating neighbor data from the user-item view. LightGCN~\cite{he2020lightgcn} simplifies the traditional graph convolution module, making it more suitable for recommendation scenarios. However, omitting modality information limits their precision in capturing user preferences.
\begin{figure*}[t]
    \centering
    \includegraphics[width=\linewidth]{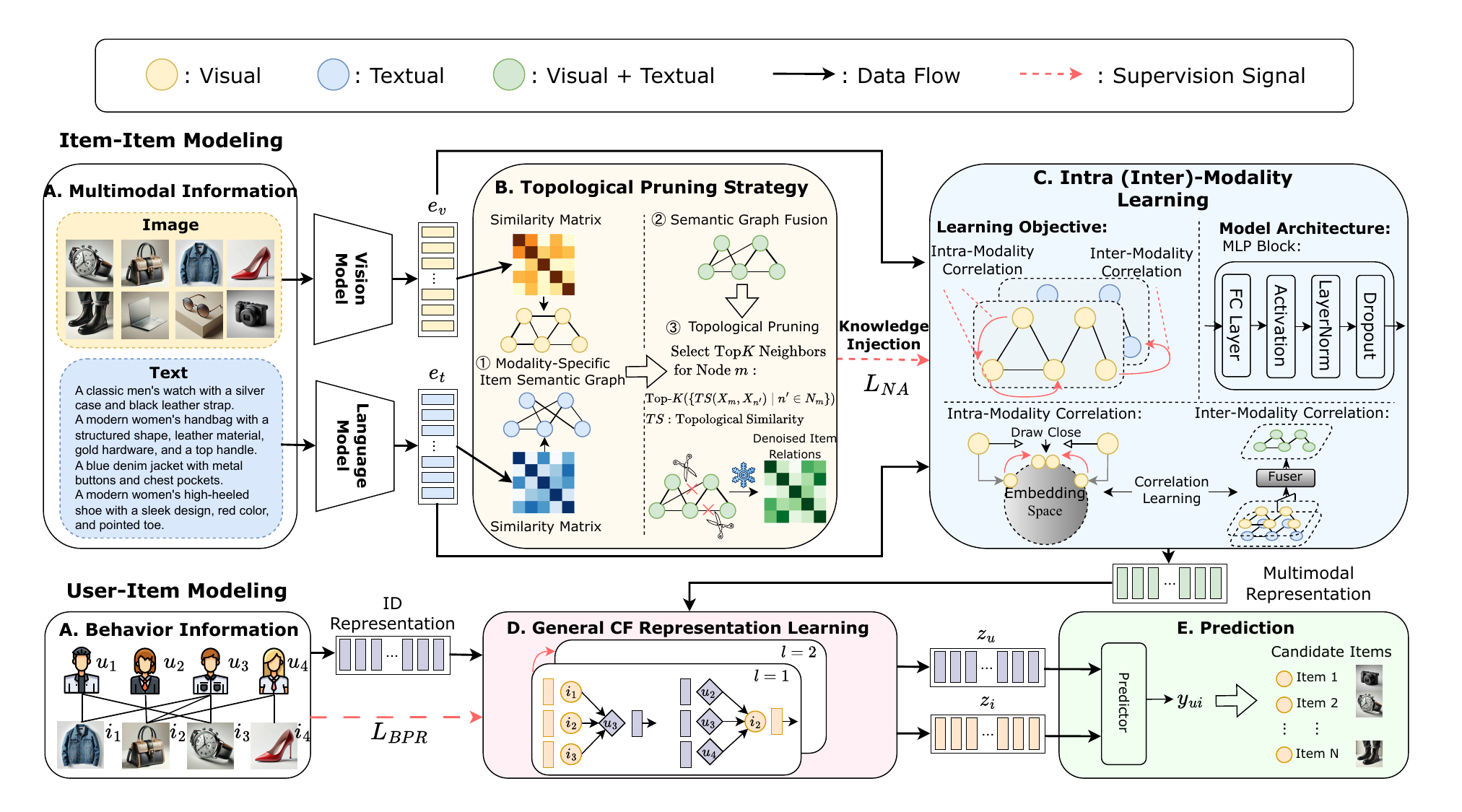}
    \caption{The framework of TMLP. It utilizes topological pruning strategy and intra (inter)-modality learning to capture modality correlations with MLPs (instead of GCNs in previous works). General CF representation learning refers to previous works that do not incorporate multimodal information, such as LightGCN.}
    \label{fig:framework}
    \vspace{-3mm}
\end{figure*}
\subsection{Multimodal Recommendation}
Multimodal recommender systems (MMRS) have evolved by integrating diverse information, such as visual and textual inputs, to alleviate data sparsity and cold-start issues in traditional CF~\cite{su2009survey}.
Early models like VBPR~\cite{he2016vbpr} combines visual features from CNNs with ID embeddings, while VECF~\cite{chen2019personalized} uses VGG~\cite{simonyan2014very} to model user attention on images and reviews. 
Later models like MMGCN~\cite{wei2019mmgcn} and GRCN~\cite{wei2020graph} enrich representations using modality-specific graphs, and 
DualGNN~\cite{wang2021dualgnn} incorporates attention mechanisms to gauge user preferences across modalities and constructs a user co-occurrence graph. 
Recently, there is a rising trend in utilizing self-supervised learning techniques within MMRS. 
SLMRec~\cite{tao2022self} integrates self-supervised tasks like feature dropping into GNNs, while
BM3~\cite{zhou2023bootstrap} employs a latent embedding dropout mechanism and  builds text and vision alignment loss.

Moreover, explicitly modeling item semantics has also been explored.
LATTICE~\cite{zhang2021mining} merges modality-specific item graphs into an integrated graph, and 
FREEDOM~\cite{zhou2023tale} enhances efficiency by discretizing and freezing the item-item graph.
MGCN~\cite{yu2023multi} further utilizes attention mechanisms to identify modality significance.
However, these approaches struggle with accurately depicting item relations due to GCN's over-smoothing issue and noise generated from directly merging modality graphs created by pre-trained models.

\subsection{Connection between MLP and GNN}
The connection between MLPs and GNNs is a fascinating area in deep learning. While MLPs process input features independently, GNNs use graph structures for message passing. Recent studies have aimed to enhance MLPs to match GNNs through techniques such as label propagation~\cite{huang2020combining}, contrastive learning~\cite{hu2021graph, dong2022node}, knowledge distillation~\cite{zhang2021graph, tian2022learning}, and additional regularization~\cite{zhang2023orthoreg}.
They seek to inject relational information into MLP, which is naturally agnostic to such information.


\section{Problem Formulation}
Denote user as $u\in \mathcal{U}$ and item as $i\in \mathcal{I}$, with $r_{u,i}$ indicating whether user $u$ has interacted with item $i$.
We construct the user rating matrix $\mathbf{R}\in \{0,1\}^{|\mathcal{U}|\times |\mathcal{I}|}$ and a user-item bipartite graph $\mathcal{G}=(\mathcal{V},\mathcal{E})$, where $\mathcal{V}$ encompasses all users and items as nodes, and $\mathcal{E}$ depicts interactions as edges. Multimodal recommendations aim to predict user preferences using the interaction graph $\mathcal{G}$ and various multimodal features.
Given user and item representations $z_u$ and $z_i$, we optimize with Bayesian Personalized Ranking (BPR)~\cite{rendle2012bpr} loss, maximizing the difference in predicted scores for interacted and non-interacted user-item pairs:
\begin{equation}\label{equ:bpr}
    \mathcal{L}_{\text{BPR}} =  \sum_{(u,i,j)}-\ln \sigma (y_{ui}-y_{uj}), \quad 
    y_{ui}=z_{u}^{T}z_{i},
\end{equation}
where, $u,i$ is a user-item pair, $j$ is a randomly sampled non-interacted item,  and $\sigma$ is the sigmoid function.
$y_{ui}$ is the predicted score, typically an inner product of $z_u$ and $z_i$.
\section{Methodology}
\subsection{Framework Overview}
Figure~\ref{fig:framework} provides an overview of TMLP, highlighting three main components: (i) Topological Pruning Strategy (TPS), which integrates and denoises item relations by leveraging the topological information of modality-specific graphs; (ii) Intra (Inter)-Modality Learning (IML), which uses MLPs to accurately model intra- and inter-modality correlations, producing comprehensive multimodal item representations by maximizing mutual information between node representations and their neighbors. TPS-derived item relations serve as additional supervision for MLP training; (iii) General Collaborative Filtering (CF) Representation Learning, which further refines user and item representations by combining multimodal and user-item interaction data, ultimately generating top-$N$ recommendations through a predictor. 

\subsection{Topological Pruning Strategy}
To ensure TMLP's robustness and avoid misguidance from noisy item relations, we propose the Topological Pruning Strategy (TPS). TPS aims to extract accurate information from pre-trained multimodal features, enhancing robustness against modality mismatches.
Following prior work~\cite{zhang2021mining}, we construct modality-specific item similarity matrices, $A_{v}$ for vision and $A_{t}$ for text, by computing feature similarity and retaining the top $K'$ edges. The latent item-item graph $A$ is then formed by a weighted sum of $A_{v}$ and $A_{t}$, with $\beta_m$ representing the importance of the visual modality. We set $K'=10$ and $\beta_m=0.1$ for consistency.
\begin{equation}\label{equ:taps1}
A = \beta_{m}\cdot A_{v} + (1-\beta_{m})\cdot A_{t}.
\end{equation}
This approach is insufficient because representations from pre-trained models are specific to each modality and may suffer from modality incompatibility, leading to coarse and noisy item relations that can mislead subsequent representation learning. Inspired by~\citet{dong2022node}, TPS refines item relations by assessing topological similarity between nodes.
Intuitively, if randomly selected nodes frequently fall within the neighborhoods of both nodes $m$ and $n$, these nodes are likely to be topologically similar.
We denote the topological information of node $m$ as $x_m$, representing either $\mathcal{N}_m$ (its neighborhood) or $\overline{\mathcal{N}}_m = \mathcal{V} - \mathcal{N}_m$ (its complement). The probability of selecting a point within its neighborhood is given by Equation~\eqref{equ:taps2}, where $a = 1$ corresponds to $\mathcal{N}_m$ and $a = 0$ corresponds to its complement. 
The topological information between nodes $m$ and $n$ is defined by the joint probability of selecting a point within both neighborhoods, as shown in Equation~\eqref{equ:taps5}:
\begin{equation}\label{equ:taps2}
\fontsize{9pt}{1pt}\selectfont
p\left(x_m = \mathcal{N}_m^a\right) = \frac{|\mathcal{N}_m^a|}{|\mathcal{V}|},
\end{equation}
\begin{equation}\label{equ:taps5}
\fontsize{9pt}{1pt}\selectfont
p\left(x_m = \mathcal{N}_m^a, x_n = \mathcal{N}_n^b\right) = \frac{|\mathcal{N}_m^a \cap \mathcal{N}_n^b|}{|\mathcal{V}|}.
\end{equation}
We define topological similarity between neighboring nodes $m$, $n$ in Equation~\eqref{equ:taps3}. Higher values indicate greater similarity. For each node $m$, we compute topological similarity with its neighbors and select the top-$K$ to represent denoised item relations. The sampling size $K$ is a tunable hyperparameter. As TPS relies solely on the topology of $A$, it can be performed before training without additional computational cost. The process is summarized in Equation~\eqref{equ:taps4}, where $\mathds{1}$ represents the indicator function and $\bar{A}$ stands for the denoised topological dependices between items:
\begin{equation}\label{equ:taps3}
\fontsize{9pt}{1pt}\selectfont
TS(X_m; X_n) = \sum_{x_m} \sum_{x_n} p(x_m, x_n) \cdot \log \frac{p(x_m, x_n)}{p(x_m) \cdot p(x_n)},
\end{equation}
\begin{equation}\label{equ:taps4}
\bar{A}_{mn} = A_{mn} \cdot 
\mathds{1}_{\{n \in \text{Top-}K(\{TS(X_m, X_{n^{'}})\}| n^{'} \in \mathcal{N}_m)\}}.
\end{equation}

\subsection{Intra-Modality Learning}
With high-quality item relations, the next challenge is making the topology-agnostic MLP aware of topological dependencies. We achieve this by maximizing mutual information~\cite{dong2022node} between node representations and their contextual neighborhoods to capture intra-modality correlations. Mutual information, a core concept in information theory, quantifies dependency between variables. By maximizing the mutual information between the representations of adjacent items, we align them to be semantically similar, which mirrors the underlying mechanics of GCNs. We denote the Probability Density Function (PDF) of the node representation $z_m$ as $p(Z(x))$ and the PDF of the contextual neighborhood representation $c_m$ as $p(C(x))$. The joint PDF is $p(C(x), Z(x))$. The mutual information between node and neighborhood representations is defined as:

\begin{equation}\label{equ:mutual}
\fontsize{8.5pt}{1pt}\selectfont
I(C(x); Z(x)) =
\int
p(C(x), Z(x)) \cdot \log \frac{p(C(x), Z(x))}{p(C(x)) \cdot p(Z(x))} dx.
\end{equation}

However, mutual information is challenging to compute, so we use Mutual Information Neural Estimation (MINE)~\cite{belghazi2018mutual} to efficiently convert this into minimizing neighborhood alignment (NA) loss.
NA loss estimates mutual information through node sampling, treating $r$-hop neighbors as contextual neighbors and other nodes as negative samples, as shown in Equation~\eqref{equ:mcloss}. The sim() function denotes cosine similarity, exp() is the exponential function, $\tau$ is the temperature parameter, and $\mathds{1}$ is the indicator function. NA loss minimizes the feature distance between a node and its contextual neighbors while maximizing the distance from negative samples, aligning node features by their connectivity for a more accurate representation space:
\begin{equation}\label{equ:mcloss}
\fontsize{6.5pt}{1pt}\selectfont
\mathcal{L}_{\text{NA}} = - \mathbb{E}_{v_m\in \mathcal{V}} \left[\log \frac{\sum_{n=1}^{B} \mathds{1}_{n \neq m} (\bar{A}_{mn})^{r} \exp (\text{sim}(z_m, z_n)/\tau)} {\sum_{k=1}^{B} \mathds{1}_{k \neq m} \exp (\text{sim}(z_m, z_k)/\tau)}\right].
\end{equation}

\subsection{Inter-Modality Learning}
We now obtain multimodal item representations, as summarized in Equation~\eqref{equ:forward1}. Our MLP structure includes linear layers followed by activation, Layernorm~\cite{ba2016layer} for stability, and dropout. $e_v$ and $e_t$ represent the multimodal features from pre-trained models.
Fuser combines these features to capture inter-modality correlations, using either a transformation matrix or simple concatenation.
After obtaining $h_{i}$, we aggregate multimodal and ID-based information with graph $\mathcal{G}$ for user-item modeling, as in Equation~\eqref{equ:lgcn1}, with $h_{u}$ randomly initialized. The aggregator can be any CF model; we use LightGCN~\cite{he2020lightgcn} for consistency with prior work.
\begin{equation}\label{equ:forward1}
\fontsize{8.5pt}{5pt}\selectfont
h_{v}=\text{MLP}_{v}(e_{v}),\quad
h_{t}=\text{MLP}_{t}(e_{t}), \quad
h_{i}=\text{Fuser}(h_{v}, h_{t}).
\end{equation}
\begin{equation}\label{equ:lgcn1}
\fontsize{9.5pt}{5pt}\selectfont
z_{u}, z_{i} =\text{Aggregator}(h_{u}, h_{i}),\
\end{equation}
user and item representations at the $(l + 1)^{th}$ GCN layer are given by Equation~\eqref{equ:lgcn2}, with $N_u$ and $N_i$ as the one-hop neighbors in graph $\mathcal{G}$.
The final representations $z_u$ and $z_i$ are obtained by summing the GCN layer outputs in Equation~\eqref{equ:lgcn3}. The prediction score, computed as the inner product of these representations in Equation~\eqref{equ:bpr}, is used to rank candidate items, with top-$N$ items recommended to users:
\begin{equation}\label{equ:lgcn2}
\fontsize{8.5pt}{5pt}\selectfont
h_u^{(l+1)} = \sum_{i \in N_u} \frac{1}{\sqrt{|N_u|}\sqrt{|N_i|}} h_i^{(l)},
\end{equation}
\begin{equation}\label{equ:lgcn3}
\fontsize{8.5pt}{5pt}\selectfont
z_u = \sum_{l=0}^{L} h_u^{(l)}, \quad z_i = \sum_{l=0}^{L} h_i^{(l)}.
\end{equation}

\subsection{Optimization Process}
Incorporating the key components mentioned above, TMLP can now capture nuanced and complex modality correlations without explicit message passing. This section outlines TMLP's optimization process, including its objective function and procedures for training and inference.
\subsubsection{Objctive Function}
To train the recommendation task, we use the BPR loss in Equation~\eqref{equ:bpr}.
Additionally, to guide TMLP with topological information, we employ an extra loss function, as shown in Equation~\eqref{equ:mcloss}. TMLP’s joint loss function is expressed in Equation~\eqref{equ:loss_fn}:
\begin{equation}\label{equ:loss_fn}
\mathcal{L}=\mathcal{L}_{\text{BPR}} + \alpha\mathcal{L}_{\text{NA}} + \lambda \|\Theta\|_2,
\end{equation}
Here, $\lambda$ is the $L_2$ regularization weight, $\Theta$ represents the model parameters, and $\alpha$ is a tunable parameter greater than zero, used to balance the losses. We train TMLP by minimizing the joint loss function on the training dataset.

\subsubsection{Training and Inference}
TPS, based on the topological structure from pre-trained modality relations, is completed before training and does not affect efficiency. During forward propagation, TMLP learns topological dependices with neighborhood alignment loss.
At inference, TMLP requires only node features without graph topological information.



\section{Experiments}~\label{sec:exp}
In this section, we present the details of our experiments.
We are interested in the following research questions:
\\
\textbf{RQ1}: How does TMLP perform against the SOTA methods?
\\
\textbf{RQ2}: How does each component affect the performance? \\
\textbf{RQ3}: What is the influence of crucial hyperparameters? \\
\textbf{RQ4}: What is the training efficiency of TMLP? \\
\textbf{RQ5}: How does TMLP handle noisy connections?

\subsection{Experiment Setup}
\subsubsection{Datasets.}
We conduct experiments on three categories from the Amazon review dataset~\cite{he2016ups,mcauley2015image}: \textit{Baby}, \textit{Sports}, and \textit{Electronics}. 
The dataset contains product descriptions and images as textual and visual features, with each review rating treated as a positive interaction.
We retain 5-core users and items, using 4096-dimensional visual features and 384-dimensional textual features from prior research~\cite{zhou2023bootstrap}.
Table~\ref{tab:dataset} summarizes the statistics for the datasets. 
\begin{table}[t]
    \centering
    \caption{Statistics of the experimental datasets.}
    \vspace{-3mm}
\scalebox{0.9}{
\begin{tabular}{ccccc}
\hline
\textbf{Dataset} & \textbf{\# Users} & \textbf{\# Items} & \textbf{\# Interactions} & \textbf{Sparsity} \\ \hline
Baby             & 19,445            & 7,050             & 160,792                 & 99.88\%           \\
Sports           & 35,598            & 18,357            & 296,337                 & 99.95\%           \\
Electronics         & 192,403            & 63,001            & 1,689,188                 & 99.99\%           \\ \hline
\end{tabular}}
\label{tab:dataset}
\vspace{-2mm}
\end{table}

\subsubsection{Baselines and Evaluation Metrics.}
We compare TMLP with several SOTA models, which can be divided into two groups: (1) General CF models: 
\textbf{BPR}~\cite{rendle2012bpr} and \textbf{LightGCN}~\cite{he2020lightgcn}.
(2) Multimodal models: 
\textbf{VBPR}~\cite{he2016vbpr}, \textbf{MMGCN}~\cite{wei2019mmgcn}, \textbf{GRCN}~\cite{wei2020graph}, \textbf{DualGNN}~\cite{wang2021dualgnn}, \textbf{BM3}~\cite{zhou2023bootstrap}, \textbf{MGCN}~\cite{yu2023multi} and \textbf{FREEDOM}~\cite{zhou2023tale}.

\begin{table*}[t]
    \centering
    \caption{Overall performance of various methods. Best results are in \textbf{boldface} and the second best is \underline{underlined}. Improv. $\uparrow$ stands for the relative improvement of TMLP over the best baseline. We conduct experiments across 5 different seeds and * indicates that the improvements are statistically significant compared of the best baseline with $p<0.01$ in t-test.}
    \vspace{-1mm}
\scalebox{0.82}{
\begin{tabular}{cc|cc|ccccccccc}
\toprule
\multirow{2}{*}{\textbf{Dataset}} & \multirow{2}{*}{\textbf{Metric}} & \multicolumn{2}{c|}{\textbf{General CF model}} & \multicolumn{9}{c}{\textbf{Multimodal model}} \\
\cmidrule(lr){3-4} \cmidrule(l){5-13}
  &  & BPR & LightGCN & VBPR & MMGCN & GRCN & DualGNN & BM3 & MGCN & FREEDOM & TMLP & Improv. $\uparrow$\\ \hline
\multirow{4}{*}{Baby} &R@10  &0.0357  &0.0479  &0.0423  &0.0378  &0.0532  &0.0448  &0.0534  &0.0620  &\underline{0.0624}  &\textbf{0.0671*} & 7.53\% \\
    &R@20  &0.0575  &0.0754  &0.0663  &0.0615  &0.0824  &0.0716  &0.0845  &0.0964  &\underline{0.0985}  &\textbf{0.1016*} & 3.15\% \\
    &N@10  &0.0192  &0.0257  &0.0223  &0.0200  &0.0282  &0.0240  &0.0285  &\underline{0.0339}  &0.0324  &\textbf{0.0360*} & 6.19\% \\
    &N@20  &0.0249  &0.0328  &0.0284  &0.0261  &0.0358  &0.0309  &0.0365  &\underline{0.0427}  &0.0416  &\textbf{0.0449*} & 5.15\% \\ 
\hline
\multirow{4}{*}{Sports} &R@10  &0.0432  &0.0569  &0.0558  &0.0370  &0.0559  &0.0568  &0.0625  &\underline{0.0729}  &0.0710  &\textbf{0.0769*} & 5.49\% \\
    &R@20  &0.0653  &0.0864  &0.0856  &0.0605  &0.0877  &0.0859  & 0.0956 &\underline{0.1106}  &0.1077  &\textbf{0.1152*} & 4.16\% \\
    &N@10  &0.0241  &0.0311  &0.0307  &0.0193  &0.0306  &0.0310  &0.0342  &\underline{0.0397}  &0.0382  &\textbf{0.0416*} & 4.79\% \\
    &N@20  &0.0298  &0.0387  &0.0384  &0.0254  &0.0389  &0.0385  &0.0427  &\underline{0.0496}  &0.0476  &\textbf{0.0515*} & 3.83\% \\ 
\hline
\multirow{4}{*}{Electronics} &R@10  &0.0235  &0.0363  &0.0293  &0.0207  &0.0349  &0.0363  & 0.0420 &\underline{0.0439}  &0.0382  &\textbf{0.0464*} & 5.69\% \\
    &R@20  &0.0367  &0.0540  &0.0458  &0.0331  &0.0529  &0.0541  & 0.0628  &\underline{0.0643}  &0.0588  &\textbf{0.0687*} & 6.84\% \\
    &N@10  &0.0127  &0.0204  &0.0159  &0.0109  &0.0195  &0.0202  & 0.0234 &\underline{0.0245}  &0.0209  &\textbf{0.0259*} & 5.71\% \\
    &N@20  &0.0161  &0.0250  &0.0202  &0.0141  &0.0241  &0.0248  & 0.0288 &\underline{0.0298}  &0.0262  &\textbf{0.0316*} & 6.04\% \\
\bottomrule
\end{tabular}}
\label{tab:overall}
\vspace{-2mm}
\end{table*}
\begin{table}[t]
    \centering
    \caption{Ablation study on different variants and modalities. 
    }
\scalebox{0.85}{
\begin{tabular}{c|c|cccc}
\toprule
Dataset & Variant & R@10 & R@20 & N@10 & N@20 \\ \hline
\multirow{7}{*}{Baby} & TMLP  &\textbf{0.0671}  &\textbf{0.1016}  &\textbf{0.0360}  &\textbf{0.0449}  \\
 &TMLP$_T$  &0.0656  &0.1010  &0.0354  &0.0445  \\
 &TMLP$_V$  &0.0538  &0.0847  &0.0296  &0.0375  \\
 &Rand Pruning  &0.0645  &0.0993  &0.0354  &0.0443  \\
 &w/o Pruning  &0.0602  &0.0926  &0.0323  &0.0406  \\
 &w/o NA Loss  &0.0519  &0.0812  &0.0281  &0.0356  \\ \hline
\multirow{7}{*}{Sports} &TMLP  &\textbf{0.0769}  &\textbf{0.1152}  &\textbf{0.0416}  &\textbf{0.0515}  \\
 &TMLP$_T$  &0.0752  &0.1128  &0.0405  &0.0502  \\
 &TMLP$_V$  &0.0644  &0.0981  &0.0349  &0.0436  \\
 &Rand Pruning  &0.0742  &0.1111  &0.0402  &0.0497  \\
 &w/o Pruning  &0.0538  &0.0838  &0.0285  &0.0362  \\
 &w/o NA Loss  &0.0617  &0.0948  &0.0337  &0.0422  \\
 \bottomrule
\end{tabular}
\label{tab:ablation}
\vspace{-3mm}
}
\end{table}
To ensure fair comparison, we use the same evaluation settings and 8:1:1 data split as in~\cite{zhou2023bootstrap} on the filtered 5-core data. 
We evaluate with NDCG@$N$ and Recall@$N$ (denoted as N@$N$ and R@$N$) for top-$N$ recommendation, averaging metrics across all test users for $N$=10 and $N$=20. During testing, we compute metrics using the all-ranking protocol based on user scores for all items.

\subsubsection{Implementation Details.}
We implement TMLP\footnote{Our code is available at 
\url{https://github.com/jessicahuang0163/TMLP}.} with PyTorch using MMRec~\cite{zhou2023mmrec}, a comprehensive repository for multimodal recommendation models.
Following~\citet{he2020lightgcn, zhang2021mining}, we set the embedding size to 64, initialized with Xavier~\cite{glorot2010understanding}.
Optimal hyperparameters are found via grid search, with learning rates in \{1e-4, 5e-4, 1e-3, 5e-3\}, MLP layers in \{2, 3, 4\}, NA loss weight $\alpha$ in Equation~\eqref{equ:loss_fn} from 0 to 2 (interval 0.1), and sampling size $K$ from 3 to 10. The regularization weight $\lambda$ is set to 0. We fix the hidden dimension of MLP as 512, activation function as ~`tanh' and dropout rate as 0. The ratio of visual features $\beta_{m}$ in Equation~\eqref{equ:taps1} is set to 0.1, with $r=1$ in Equation~\eqref{equ:mcloss}.
Training is capped at 1000 epochs with early stopping at 20 consecutive epochs.
We select the best models based on $Recall@20$ on the validation set and then report metrics on the test set.
\subsection{Performance Comparison(RQ1)}
Table~\ref{tab:overall} reports the recommendation performance of different models in terms of Recall and NDCG on three datasets, from which we have the following observations:
\begin{itemize}[leftmargin=*]
    \item TMLP outperforms existing multimodal baselines across all three datasets. 
    Specifically, TMLP improves upon the second-best model by $7.53\%$, $5.49\%$, and $5.69\%$ in $Recall@10$.
    This improvement is due to our use of MLPs instead of shallow GCNs to better capture high-dimensional multimodal information and avoid over-smoothing. Additionally, topological pruning strategy (TPS) denoises item relations from pre-trained visual and textual models, leading to more robust representations.
    \item Some baselines like FREEDOM, perform well on \textit{Baby} and \textit{Sports}, but underperform in \textit{Electronics}, which has a larger user and item base. This is because FREEDOM constructs the item-item graph using multimodal information but only considers ID embeddings in recommendation process, neglecting multimodal integration. In contrast, TMLP integrates visual, textual, and ID information comprehensively, and bridges modality correlations by intra (inter)-modality  learning. Additionally, TMLP's adjustable MLP architecture adapts to both small and large datasets, achieving better performance.
    \item The results show that models leveraging multimodal information consistently outperform matrix factorization (e.g., BPR) and graph-based models (e.g., LightGCN) that rely solely on ID features. This underscores the importance of incorporating multimodal knowledge and advanced correlation modeling in recommendations.
\end{itemize}

\subsection{Ablation Study (RQ2)}
In this section, we conduct extensive experiments to analyze the performance of TMLP under different settings. The results on \textit{Baby} and \textit{Sports} are shown in Table~\ref{tab:ablation}.
\subsubsection{Effect of Different Components of TMLP}
To examine the role of various components, we design the following variations:
(i) \textbf{w/o NA Loss} disables correlation modeling by setting $\alpha =0$ in Equation~\eqref{equ:loss_fn}, and uses a vanilla MLP to capture item relations;
(ii) \textbf{w/o Pruning} excludes TPS, relying on original item relations from FREEDOM for supervising correlation learning;
(iii) \textbf{Rand Pruning} randomly samples the same number of neighbors as TMLP.

From Table~\ref{tab:ablation}, we observe:
(i) NA loss is crucial in TMLP, significantly enhancing the model's ability to learn inter- and intra-modality correlations. Without it, an MLP fails to capture these relationships. (ii) Topological Pruning Strategy (TPS) is also essential. While acceptable on the smaller \textit{Baby} dataset, omitting it on the larger \textit{Sports} dataset leads to significant performance degradation due to increased noise. (iii) We find that random pruning performs slightly better due to small sample size ($K=5$ in \textit{Sports}) filtering out noise. However, the gap between random pruning and TMLP highlights the strength of TPS.

\subsubsection{Effect of Single versus Multi-Modalities}
We compare TMLP's performance in uni-modal and multimodal settings. $TMLP_V$, $TMLP_T$ and $TMLP$ represent models using visual, textual, and both modalities.
Table~\ref{tab:ablation} indicates that both modalities are indispensable, while Amazon's textual features contribute more to performance than visual features.

\subsection{Hyperparameter Sensitivity Study (RQ3)}
\subsubsection{MLP-related Hyperparameters}
\begin{figure}[ht]
    \centering
    \includegraphics[width=\linewidth]{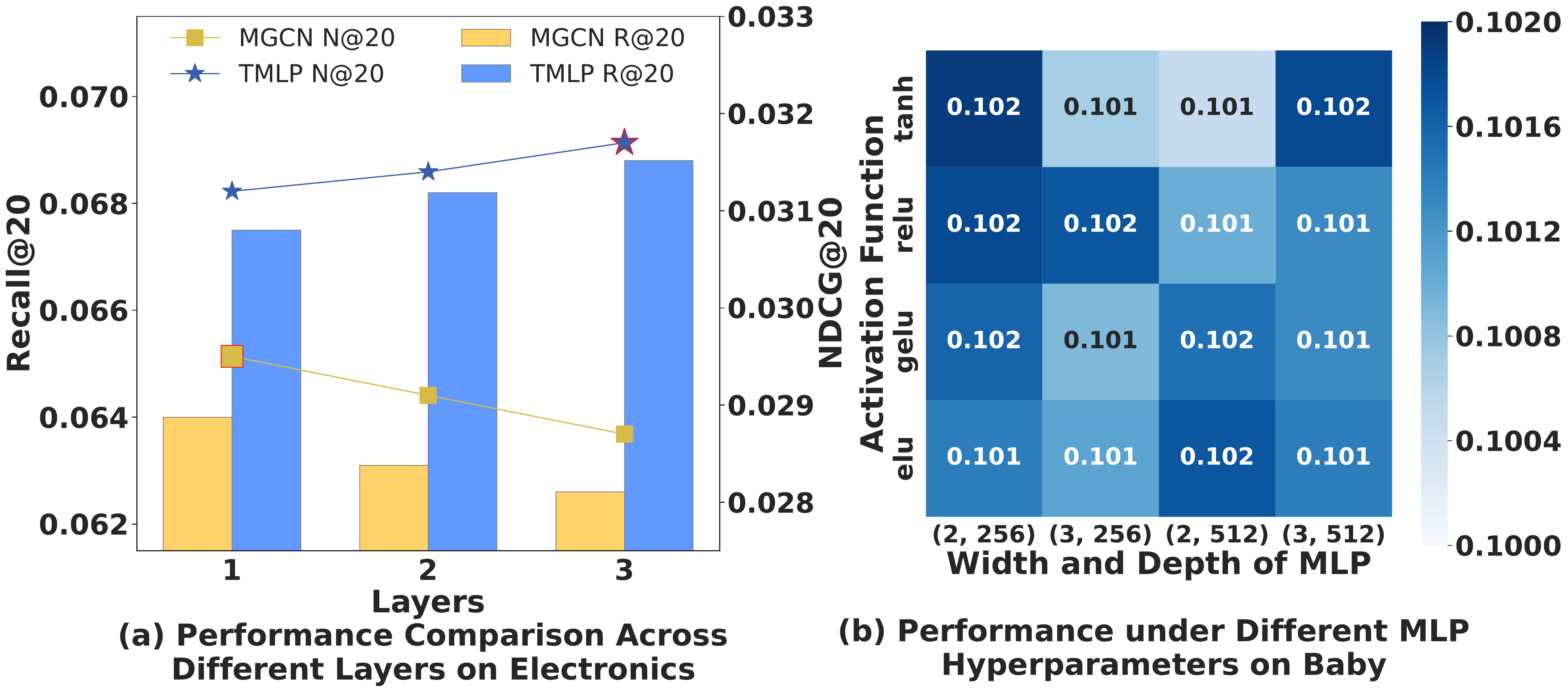}
    \caption{Hyperparameter study on width, depth and activation functions of MLP on \textit{Baby} and \textit{Electronics}.}
    \vspace{-3mm}
    \label{fig:mlp}
\end{figure}

We explore the impact of MLP-related hyperparameters like width, depth, and activation functions.
Figure~\ref{fig:mlp}(a) shows that on larger datasets like \textit{Electronics}, TMLP's performance improves with greater depth.
In Figure~\ref{fig:mlp}(b), TMLP shows stable Recall@20 on smaller datasets like \textit{Baby}, indicating its robustness to these parameters.
This highlights TMLP's effectiveness in capturing complex cross-modality correlations.

\subsubsection{Number of NA Loss Weight $\alpha$}
\begin{figure}[ht]
    \centering
    \includegraphics[width=\linewidth]{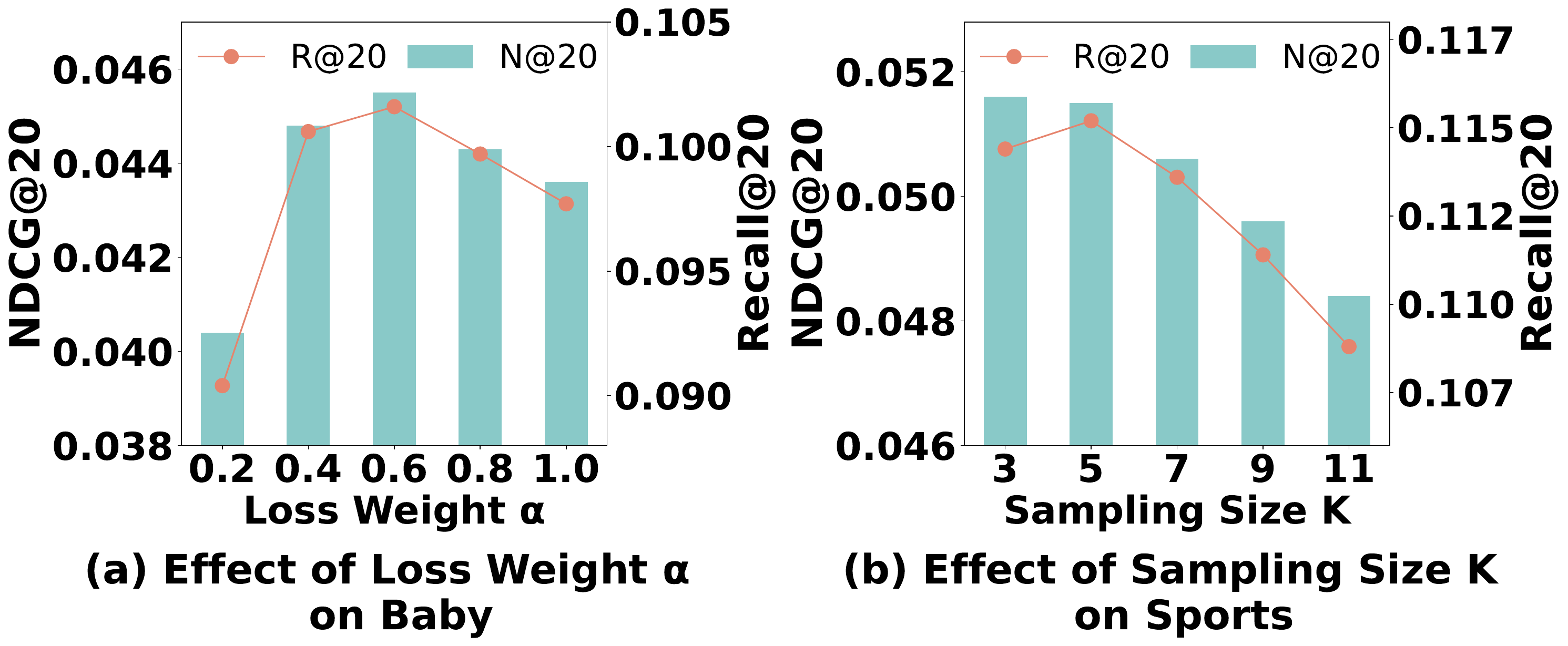}
    \caption{Hyperparameter study on NA loss weight $\alpha$ and sampling size $K$ on \textit{Baby} and \textit{Sports}.}
    \vspace{-5mm}
    \label{fig:hyper}
\end{figure}

We study the effect of NA loss weight $\alpha$ on \textit{Baby} by tracking changes in Recall@20 and NDCG@20 in Figure~\ref{fig:hyper}(a).
The metrics initially increase and then decrease, indicating an optimal $\alpha$. A small $\alpha$ underfits inter- and intra-modality correlations, while a large $\alpha$ neglects the BPR loss, harming performance. 

\subsubsection{Number of Sampling Size $K$}
We examine the effect of sampling size $K$ on \textit{Sports} by tracking changes in Recall@20 and NDCG@20 in Figure~\ref{fig:hyper}(b).
With an average node degree of 17-18, TMLP performs best at $K=5$. A smaller $K$ slightly reduces performance due to insufficient neighbors capturing modality correlations, yet TMLP still outperforms other models. Conversely, increasing $K$ introduces noise, leading to a sharp performance drop.

\subsection{Training Efficiency (RQ4)}
\begin{figure}[ht]
    \centering
    \includegraphics[width=\linewidth]{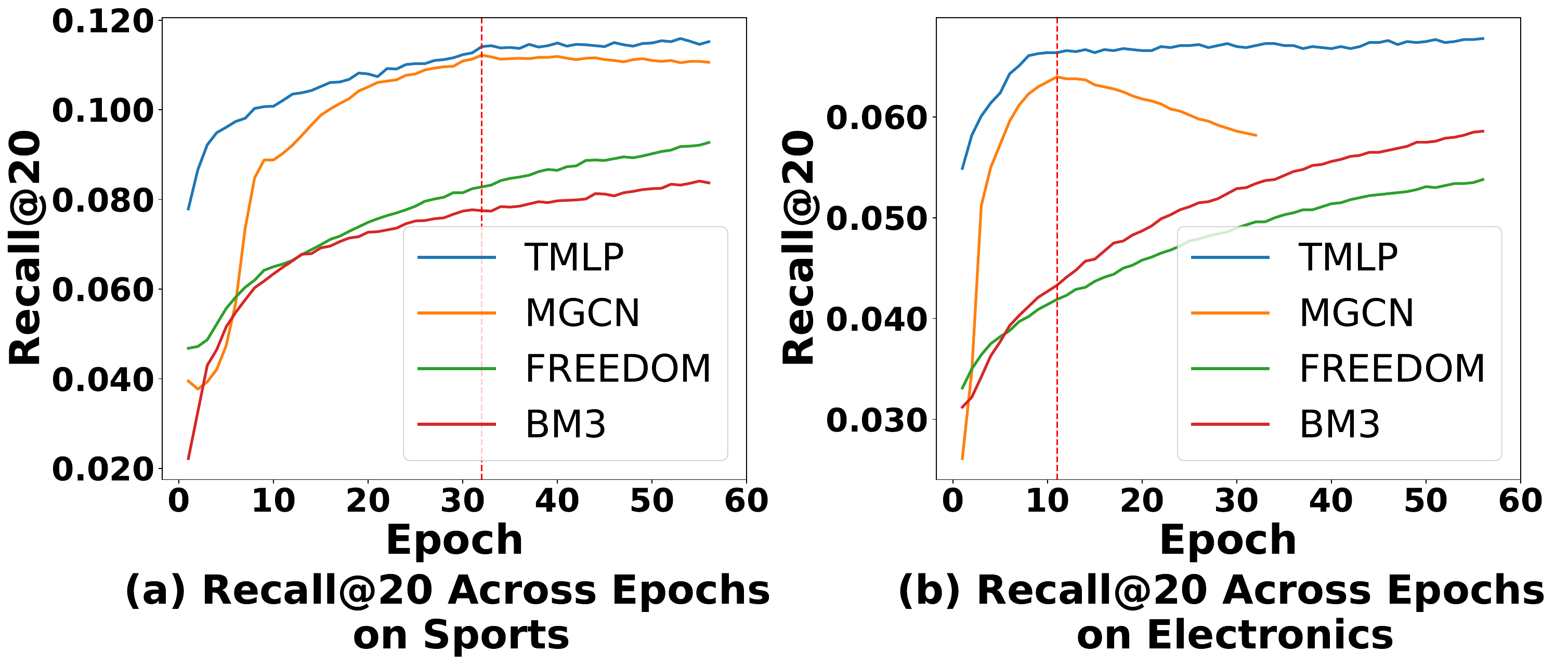}
    \caption{Comparison of training efficiency.}
    \vspace{-5mm}
    \label{fig:efficiency}
\end{figure}

We compare TMLP's efficiency with other models. 
Figure~\ref{fig:efficiency} shows Recall@20 across epochs on \textit{Sports} and \textit{Electronics}.
With consistent hyperparameters (learning rate: 0.001, batch size: 2048), TMLP converges faster, reaching peak performance by the 30$^{th}$ on \textit{Sports} and 10$^{th}$ epoch on \textit{Electronics}. In contrast, FREEDOM and BM3 take around 200 epochs and still underperform. MGCN converges quickly but experiences a sharp performance drop, triggering early stopping.
Notably, TMLP also has a shorter time per epoch compared to existing SOTA models, as shown in Figure~\ref{fig:intro}(a).
Additionally, TMLP's performance curve is more stable and smooth.

\subsection{Resilience to Corrupted Data (RQ5)}
In real world, mismatches and misalignments often introduce noise in item adjacency matrix $A$, which we refer to as corrupted connections.
Our aim is for multimodal models to still learn effective cross-modality correlations in these scenarios.
We define the corruption ratio $\epsilon$ as the proportion of noise, with the corrupted relations $A^{'}$ expressed as:
\begin{equation}\label{equ:robust1}
    A^{'} = P \odot A + (1 - P) \odot N
\end{equation}
\begin{equation}\label{equ:robust2}
P_{ij}= \begin{cases}0, & \text{probability=}\epsilon \\ 1, & \text {probability=}1-\epsilon\end{cases}
\end{equation}
\begin{figure}[ht]
    \centering
    \includegraphics[width=\linewidth]{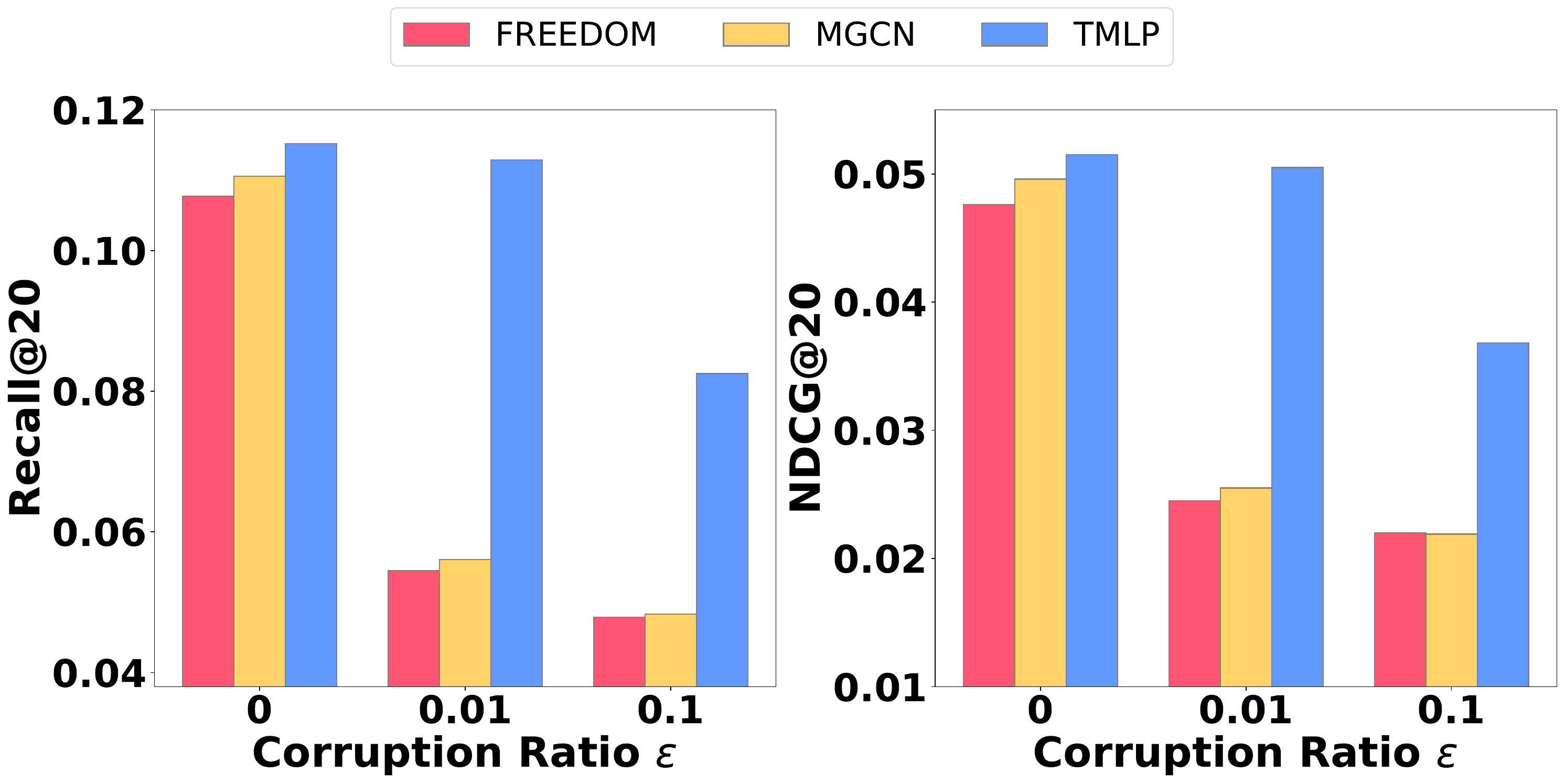}
    \caption{Performance on corrupted data on Sports.}
    \vspace{-2mm}
    \label{fig:robust}
\end{figure}
\( A \) is the original item adjacency matrix and \( N \) represents random noise.
The operation \( \odot \) indicates element-wise multiplication and 
\( P \) is a binary matrix determined by Equation~\eqref{equ:robust2}.
In Figure~\ref{fig:robust}, TMLP shows strong performance on \textit{Sports} despite noise. With a 1\% noise ratio, TMLP effectively filters out irrelevant information via its topological pruning strategy, maintaining performance. In contrast, models like FREEDOM and MGCN suffer significant performance drops with even the slightest disturbance. These models even fall behind LightGCN, showing that multimodal integration with minor noise severely impacts recommendation quality, highlighting their lack of robustness. 

\subsection{Case Study}
\begin{figure}[ht]
    \centering
    \includegraphics[width=1.02\linewidth]{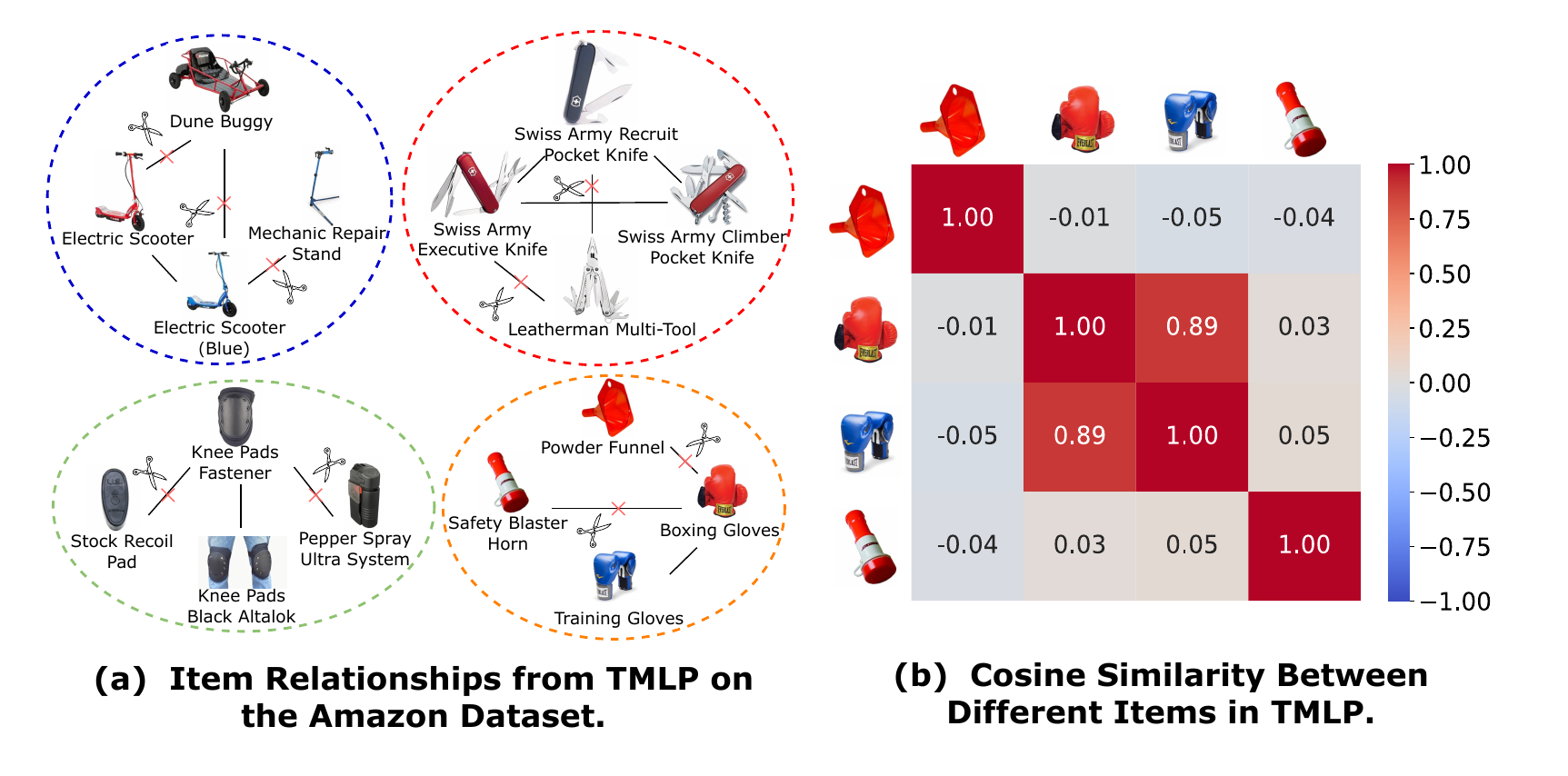}
    \caption{Case study on TMLP's discriminative power.}
    \label{fig:case}
\end{figure}

We conduct a case study to evaluate TPS and demonstrate TMLP's discriminative ability. In Figure~\ref{fig:case}(a), clustered circles represent items connected in the adjacency matrix of existing MMRS. TPS prunes these connections, retaining truly similar items while removing links between unrelated ones. Figure~\ref{fig:case}(b) presents a heatmap of cosine similarity in TMLP, illustrating that TMLP captures the similarity between boxing gloves while distancing other items, highlighting its strong discriminative ability.
\section{Conclusion}
In this paper, we introduce TMLP, a novel topology-aware MLP framework that enhances scalability by replacing GCNs in MMRS.
TMLP captures high-order and complex multimodal correlations through intra (inter)-modality learning, which injects topological dependencies between items into MLPs.
To mitigate the impact of noisy connections that could degrade performance, TMLP employs a topological pruning strategy to effectively denoise item relations, improving robustness against modality mismatches.
Our evaluation on three real-world datasets shows that TMLP outperforms nine baseline models, demonstrating its effectiveness, superior training efficiency and robustness.

\section*{Acknowledgements}
This work is partially supported by Shanghai Municipal Science and Technology Major Project (2021SHZDZX0102) and National Natural Science Foundation of China (62322603, 62177033).

\bibliography{TMLP}

\clearpage

\section*{Appendix}

\subsection*{Compared Baselines}

\begin{itemize}
\item \textbf{BPR}~\cite{rendle2012bpr}: This model captures latent factors representing both users and items from historical user-item interactions. It is known for introducing the widely adopted BPR loss function.
\item \textbf{LightGCN}~\cite{he2020lightgcn}:
LightGCN is a popular GCN-based collaborative filtering method that streamlines GCN design for improved recommendation performance. 
\item \textbf{VBPR}~\cite{he2016vbpr}:
VBPR integrates visual features with ID embeddings to form item representations. It utilizes the BPR loss to capture user preferences.
\item \textbf{MMGCN}~\cite{wei2019mmgcn}:
The model discerns modality-specific preferences by creating distinct graphs for each modality and employing these user-item graphs to learn representations within each modality. Finally, the model aggregates all modality-specific representations to get the final user and item embeddings. 
\item \textbf{GRCN}~\cite{wei2020graph}:
This model employs the same user-item graphs as prior GCN-based models, but it identifies and eliminates false-positive edges within the graph. The refined graph is subsequently used to generate new representations through aggregation and information propagation.
\item \textbf{DualGNN}~\cite{wang2021dualgnn}:
This model introduces a novel user-user co-occurrence graph by utilizing representations derived from modality-specific graphs and integrating neighbor representations.
\item 
\textbf{BM3}~\cite{zhou2023bootstrap}: This method streamlines the self-supervised approach by eliminating the need for randomly sampled negative examples. It uses the dropout technique to generate contrastive views and applies three contrastive loss functions to optimize the resulting representations.
\item \textbf{MGCN}~\cite{yu2023multi}:
This model addresses modality noise by refining modality features using item behavior information. It aggregates multi-view representations to enhance recommendation performance.
\item \textbf{FREEDOM}~\cite{zhou2023tale}: This model uncovers latent structures between items by constructing an item-item graph and freezing the graph before training. It introduces degree-sensitive edge pruning techniques to reduce noise in the user-item interaction graph.
\end{itemize}

\begin{table}[t]
    \centering
    \caption{Detailed hyperparameters when training TMLP.}
    \scalebox{0.85}{
    \begin{tabular}{l|c}
    \toprule
       Hyperparameters  & Values \\
    \midrule
    Embedding dimension &64\\
    MLP hidden size & 512 \\
    MLP layers & \{2,3,4\} \\
    LightGCN layers on interaction graph& 2\\
    kNN $K'$ when fusing $A_{v}$ and $A_{t}$ & 10 \\
    $\beta_{m}$ when fusing $A_{v}$ and $A_{t}$ & 0.1 \\
    \midrule
       Learning rate  & \{1e-4, 5e-4, 1e-3, 5e-3\} \\
       Batch size & 2048 \\
       Optimizer & Adam \\
       NA loss weight $\alpha$ & 0 to 2 (interval 0.1) \\
       Temperature in NA loss $\tau$ & 1\\
       $r$-hop neighbors $r$ & 1 \\
       $L_2$ regularizer weight $\lambda$ &\{1e-3, 1e-2, 1e-1, 0\} \\
    \bottomrule
    \end{tabular}
    }
    \label{tab:appendix-hyper}
\end{table}

\begin{table*}[t!]
\centering
\caption{Ablation study with mean and std values.}
\begin{tabular}{c|c|cccc}
\toprule
\textbf{Dataset} & \textbf{Variant} & \textbf{R@10} & \textbf{R@20} & \textbf{N@10} & \textbf{N@20} \\
\midrule
\multirow{6}{*}{\textbf{Baby}} 
 & TMLP       & \textbf{0.0671$\pm$0.0011} & \textbf{0.1016$\pm$0.0008} & \textbf{0.0360$\pm$0.0005} & \textbf{0.0449$\pm$0.0004} \\
 & TMLP$_T$        & 0.0656$\pm$0.0003          & 0.1010$\pm$0.0007          & 0.0354$\pm$0.0003          & 0.0445$\pm$0.0002 \\
 & TMLP$_V$        & 0.0538$\pm$0.0009          & 0.0847$\pm$0.0006          & 0.0296$\pm$0.0006          & 0.0375$\pm$0.0003 \\
 & Rand Pruning        & 0.0645$\pm$0.0006          & 0.0993$\pm$0.0009          & 0.0354$\pm$0.0002          & 0.0443$\pm$0.0002 \\
 & w/o Pruning     & 0.0602$\pm$0.0010          & 0.0926$\pm$0.0007          & 0.0323$\pm$0.0004          & 0.0406$\pm$0.0003 \\
 & w/o NA Loss       & 0.0519$\pm$0.0004          & 0.0812$\pm$0.0012          & 0.0281$\pm$0.0005          & 0.0356$\pm$0.0006 \\
\midrule
\multirow{6}{*}{\textbf{Sports}} 
 & TMLP       & \textbf{0.0769$\pm$0.0004} & \textbf{0.1152$\pm$0.0005} & \textbf{0.0416$\pm$0.0001} & \textbf{0.0515$\pm$0.0001} \\
 & TMLP$_T$        & 0.0752$\pm$0.0006          & 0.1128$\pm$0.0006          & 0.0405$\pm$0.0003          & 0.0502$\pm$0.0003 \\
 & TMLP$_V$        & 0.0644$\pm$0.0003          & 0.0981$\pm$0.0002          & 0.0349$\pm$0.0001          & 0.0436$\pm$0.0001 \\
 & Rand Pruning        & 0.0742$\pm$0.0005          & 0.1111$\pm$0.0005          & 0.0402$\pm$0.0003          & 0.0497$\pm$0.0003 \\
 & w/o Pruning     & 0.0538$\pm$0.0007          & 0.0838$\pm$0.0008          & 0.0285$\pm$0.0004          & 0.0362$\pm$0.0004 \\
 & w/o NA Loss       & 0.0617$\pm$0.0005          & 0.0948$\pm$0.0008          & 0.0337$\pm$0.0004          & 0.0422$\pm$0.0005 \\
\bottomrule
\end{tabular}
\label{tab:appendix-ablation}
\end{table*}

\subsection*{Hyperparameter Settings}
During the training process of TMLP, the specific hyperparameters are detailed in Table~\ref{tab:appendix-hyper}, with some basic settings adapted from MMRec~\cite{zhou2023mmrec}.

\subsection*{Mean and STD Values in Ablation Study}
To further verify the effectiveness of our framework, we conduct ablation studies using five different seeds, and present the mean and standard deviation values in Table~\ref{tab:appendix-ablation}. The results demonstrate that each component contributes to the performance of TMLP.

\subsection*{Further Hyperparameter Analysis}
\begin{figure}[ht]
    \centering
    \includegraphics[width=\linewidth]{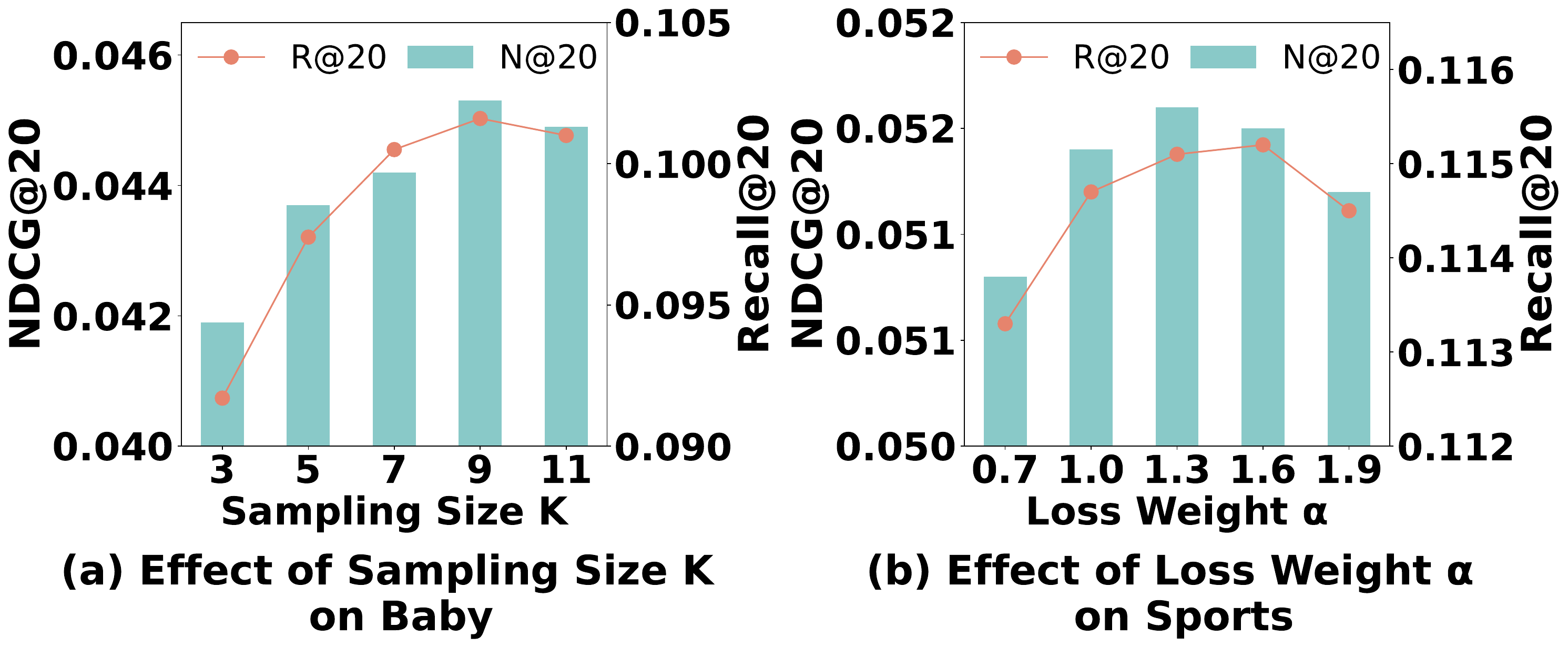}
    \caption{Hyperparameter study on NA loss weight $\alpha$ and sampling size $K$ on \textit{Baby} and \textit{Sports}.}
    \label{fig:hyper2}
\end{figure}

Building on Section 5 in the original paper, Figure~\ref{fig:hyper2} presents additional hyperparameter analysis across various datasets, examining the effect of sampling size $K$ on \textit{Baby} and NA loss weight $\alpha$ on \textit{Sports}.

\subsection*{Visualization Analysis}
\begin{figure}[ht]
    \centering
    \includegraphics[width=\linewidth]{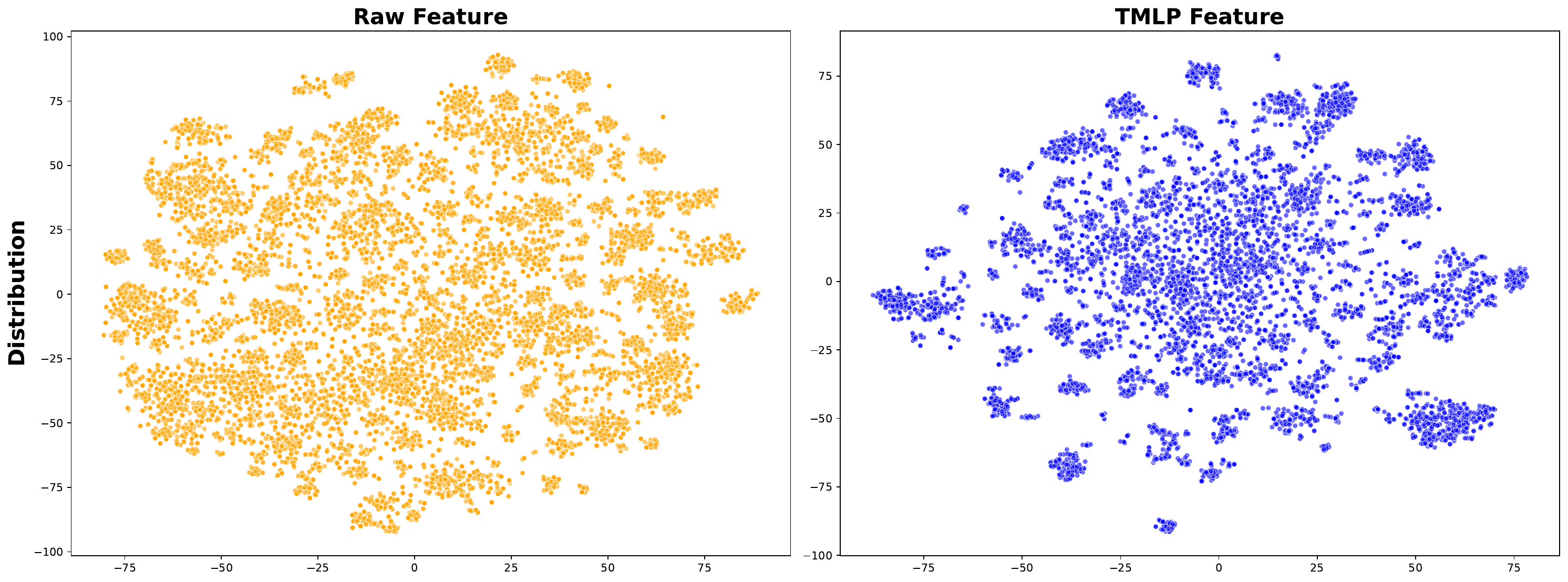}
    \caption{Item representation distribution. The left side of the figure presents the t-SNE visualization of raw features, while the right side showcases the features generated by TMLP.}
    \label{fig:tsne}
\end{figure}
The t-SNE visualization on the left in Figure~\ref{fig:tsne} illustrates the distribution of raw features, where data points are uniformly spread, forming a homogeneous cluster. This lack of distinct clusters suggests limited discriminative power, which is crucial for accurate item differentiation in recommendations.
Conversely, the t-SNE visualization on the right, representing features generated by TMLP, shows well-defined clusters. This indicates that TMLP enhances item representation by effectively capturing underlying relationships and distinctions, thereby improving the model's ability to differentiate between items for personalized recommendations.

\subsection*{Runtime Efficiency}
\begin{table}[h!]
\centering
\scalebox{0.78}{
\begin{tabular}{c|c|c|c}
\hline
\textbf{Model} & \textbf{Epochs} & \textbf{Total Training Cost} & \textbf{Performance (R@20)} \\ \hline
BM3           & 264            & 32 min                      & 0.0956                     \\ \hline
FREEDOM       & 260            & 32.3 min                    & 0.1077                     \\ \hline
MGCN          & 64             & 8.9 min                     & 0.1106                     \\ \hline
TMLP          & 73             & 10.2 min                    & 0.1152                     \\ \hline
\end{tabular}
}
\caption{Efficiency comparison across models.}
\label{tab:model_comparison}
\end{table}
As noted in Section 4.5, topological pruning strategy (TPS) relies on pre-trained modality structures and is completed before training, so it does not impact efficiency. During forward propagation, we employ MLPs and refer to topological dependencies only in calculating the NA loss, allowing flexibility. At inference time, TMLP requires only node features without graph topology, achieving higher performance and efficiency than GCN-based methods. Efficiency based on convergence speed is shown in Figure~\ref{fig:efficiency}, with total training cost included in Table~\ref{tab:model_comparison}.

\end{document}